\documentclass{article}
\usepackage{arxiv}
\usepackage{siunitx}
\usepackage{moreverb}

\usepackage[colorlinks,bookmarksopen,bookmarksnumbered,citecolor=red,urlcolor=red]{hyperref}
\usepackage{subfigure}
\usepackage{amsmath}
\usepackage{xcolor}
\usepackage{color, soul}
\usepackage{array}
\usepackage{tabularx}
\usepackage{adjustbox}
\usepackage{multirow}
\usepackage{tikz}
\usepackage{changepage}

\usepackage{natbib}
\usepackage[utf8]{inputenc} 
\usepackage[T1]{fontenc}    
\usepackage{url}            
\usepackage{booktabs}       
\usepackage{amsfonts}       
\usepackage{nicefrac}       
\usepackage{microtype}      
\usepackage{lipsum}
\usepackage{graphicx}

\title{\large Estimating the Complier Average Causal Effect in Randomised
Controlled Trials with Non-Compliance: A Comparative
Simulation Study of the Instrumental Variables and Per-Protocol
Analyses}

\author{
 Theodosios Papazoglou \\
  Department of Mathematics\\
  Imperial College of London\\
  South Kensington, SW7 2AZ \\
  \texttt{theo.papazoglou23@imperial.ac.uk} \\
   \And
 Ed Waddingham \\
  School of Public Health\\
  Imperial College of London\\
  White City, W12 0BZ \\
  \texttt{e.waddingham@imperial.ac.uk} \\
  \And
 Alastair Young \\
  Department of Mathematics\\
  Imperial College of London\\
  South Kensington, SW7 2AZ \\
  \texttt{alastair.young@imperial.ac.uk} \\
}

\begin{document}
\maketitle
\begin{abstract}
\textbf{Objective:} Randomised controlled trials (RCTs) are widely considered as gold standard for assessing the effectiveness of new health interventions. When treatment non-compliance is present in RCTs, the treatment effect in the subgroup of participants who complied with their original treatment allocation, the Complier Average Causal Effect (CACE), is a more representative measure of treatment efficacy than the average treatment effect. Through simulation we aim to compare the two most common methods employed in practice to estimate CACE. \textbf{Methods:} We considered the Per-Protocol and Instrumental Variables (IV) analyses. Based on a real study, we simulated hypothetical trials by varying factors related to non-compliance and compared the two methods by the bias of the estimate, mean squared error and $95\%$ coverage of the true value. \textbf{Results:} For binary compliance, the IV estimator was always unbiased for CACE, while the Per-Protocol estimator was unbiased for random non-compliance or when participants with good or bad conditions always received the treatment. For partial compliance, the IV estimator was less biased when participants with better conditions always received the treatment and those with worse conditions always received the control or vice versa, while the Per-Protocol estimator was less biased when participants with good or bad conditions never received the treatment.
\end{abstract}


\section{Introduction}
Randomised controlled trials (RCTs) are widely considered as gold standard for evaluating the effectiveness of an intervention and can establish a causal relationship between the intervention and an improved disease outcome \citep{Tripepietal.2020}. In RCTs, participants are randomly assigned to either a treatment or a control group, where in the latter, participants may either receive an existing standard treatment or a placebo. The statistical framework for causal inference in RCTs is based on the potential outcomes framework which was first introduced by \citet{Neyman1923} for randomised experiments and was later extended by \citet{Rubin1974} for randomised and non-randomised experiments, hence it is usually referred to as Rubin's Causal Model (RCM). In this framework, each participant is assumed to have a pair of potential outcomes which are sometimes called counterfactual outcomes; an outcome that would have been observed under the active treatment and an outcome that would have been observed under the control treatment. A causal effect, for an individual, is defined as the difference between these two outcomes. In reality, however, only one of the two outcomes can be observed and thus causal inference cannot be made at an individual level. Instead, a causal effect can be defined at the population level by considering the difference of the average value of the potential outcomes when all participants are assigned to the active treatment and the average value of the potential outcomes when all participants are assigned to the control treatment. This is usually referred to as the Average Treatment Effect (ATE) and is the primary estimand of interest in an RCT.

The main analysis employed to estimate the ATE is an Intention-to-Treat (ITT) analysis. The ITT analysis compares the outcomes of participants based on their treatment assignment regardless of the actual receipt of treatment \citep{Tripepietal.2020}. However, in RCTs, treatment compliance is rarely perfect \citep{Kamper2021}. Common examples of non-compliance are treatment crossover, failure to comply with the original treatment allocation and administrative error to name a few. The presence of non-compliance makes inference for causal effects more complicated \citep{LittleRubin2000}. In the presence of non-compliance, the ATE is a representative measure of treatment effectiveness and not of treatment efficacy; the effect of actually receiving the treatment, which might be of interest to investigators \citep{McNamee2009}. A more meaningful measure of treatment efficacy is the treatment effect in the subgroup of participants who complied with the original treatment allocation, which is called the Complier Average Causal Effect (CACE) or the Local Average Treatment Effect (LATE) and was first introduced by \citet{ImbensAngrist1994} and later extended by  \citet{Angrist1996}. Imbens and Angrist stratified participants into four different strata based on their potential treatment receipt under assignment of treatment and control. The four different strata are the compliers, who would receive the treatment if assigned to it and would not receive the treatment if assigned to the control, the never-takers, who would never receive the treatment regardless of their assignment, the always-takers, who would always receive the treatment regardless of assignment and the defiers, who would always do the opposite of what they were assigned to. In practice, we can only observe the treatment receipt for one of the two assigned treatments. The stratification into the four different strata based on the potential receipt of treatment creates a conceptual definition of compliers that is different to the observed compliers and should not be confused. Specifically, we can only observe never-takers in the treatment group but not in the control group and similarly, we can observe always-takers in the control group but not in the treatment group. Provided that certain assumptions hold \citep{Angrist1996}, we can estimate the proportions of true compliers, never-takers and always-takers using the observed data. Additionally, under these assumptions, CACE is a well-defined causal effect and is an example of a principal stratum estimand, as it compares the potential outcomes under standard versus new treatment within the same subgroup of participants, the compliers stratum \citep{Frangakis2002Principal-Strat}.

Two methods are usually employed in practice to estimate CACE, the Per-Protocol analysis and the Instrumental Variables (IV) analysis; although the Per-Protocol analysis is usually criticised in the literature as a method that can estimate neither CACE nor any other meaningful causal effect \citep{LittleRubin2000}$,$ \citep{BangDavis2007}. A Per-Protocol analysis excludes participants who did not comply with the study protocol and compares participants who adopted their original assigned treatments based on the study protocol. By doing so, participants in the Per-Protocol analysis may no longer be comparable for known or unknown confounding characteristics that are related to compliance and results tend to be biased \citep{Kamper2021}. An alternative method is the IV analysis which was first introduced as a method to estimate CACE in \citep{ImbensAngrist1994} and extended in \citep{Angrist1996}, \citep{ImbensRubin1997} and  \citep{LittleandYau1998}. An IV analysis, uses the treatment assignment as an instrument and estimates CACE using a two-stage least squares regression. The IV method requires strong assumptions hold which are not always easily verifiable in practice, however it overcomes the main limitation of the Per-Protocol analysis by preserving randomisation therefore it is expected to provide unbiased results. 

In the past, several studies have employed simulation analyses in order to compare statistical methods that account for non-compliance in RCTs \citep{Abeletal.2023}. Specifically, they have considered a wide range of trials by varying multiple factors that are associated with non-compliance such as the randomness of non-compliance, the degree of non-compliance and the type of non-compliers to name a few, and have provided valuable insight in the performance of existing methods to estimate the ATE; among which the Per-Protocol and the IV analyses. However, as far as we are concerned, no study in the past has conducted any simulation analysis in order to compare the performance of the two methods in estimating CACE. Furthermore, only one study conducted by \citet{BangDavis2007}, has considered a simulation design which incorporates the concept of the potential treatment receipt and hence the true compliance behavior as defined by \citet{ImbensAngrist1994}. Instead, past studies have focused on the observed compliance behavior since their primary endpoint was estimation of the ATE and not CACE. The simulations we develop follow the stratification framework of \citep{Angrist1996} and the primary estimand of interest is CACE and not ATE. 

In this paper, we try to provide insight in the performance of the Per-Protocol and the IV analyses in estimating both the ATE and CACE by employing multiple simulation analyses. Specifically, we simulate a wide range of trials by varying a wide range of factors associated with non-compliance. Our simulations are based on an existing clinical trial of the Continuous Positive Airway Pressure (CPAP) treatment in women with high-risk pregnancy and obstructive sleep apnea with primary endpoint the modulation of blood pressure \citep{Tantrakuletal.2023}. The simulation procedures we employ closely follow the work conducted in \citep{BangDavis2007} and  \citep{Yeetal.2014} but extend it further by exploring additional non-compliant scenarios and consider a different estimand (CACE) to be of primary interest. Additionally, we consider multiple sensitivity analyses on the majority of the parameters related to non-compliance in order to allow for greater generalisation of our findings. Our objective is to create a guide for fellow statisticians and clinical researchers on which of the two methods is more appropriate for estimating CACE depending on the various non-compliant conditions we consider that are likely to appear in real RCTs.

In Section \ref{sec:methods}, we provide a concise description of the CACE framework as introduced in \citep{ImbensAngrist1994} and define mathematically the Per-Protocol and IV methods. Additionally, we introduce the CPAP study based on which we develop our simulations and describe the simulation framework we employ. In Section \ref{sec:results}, we provide the results from our analyses and finally in Section \ref{sec:discussion}, we provide a general discussion on the results obtained from our simulations along with the strengths and limitations of our study.

\section{Methods}\label{sec:methods}
In this section we begin by providing a short but detailed description of the CACE framework as introduced in  \citep{ImbensAngrist1994}. Then, we present the two methods for estimating CACE, the Per-Protocol and the IV analyses. Finally, we describe the CPAP study we adapt and present our simulation framework.

\subsection{The CACE Framework}\label{subsec:CACE}
We assume that participants are randomly assigned to either treatment or control. Let $Z_i$ be the treatment assignment indicator, where $Z_i=1$ corresponds to the active treatment and $Z_i=0$ corresponds to the control treatment, and let $D_i(Z_i)$ be a binary indicator for whether the $i-$th participant would receive the treatment given the randomly allocated treatment, where $1$ corresponds to receipt of the active treatment and $0$ to non-receipt of the active treatment. In practice, we observe only the actual receipt of treatment $D_i$, \begin{equation}\label{eq:D}
    D_i=Z_iD_i(1)+(1-Z_i)D_i(0).
\end{equation}
Additionally, let $Y_i(Z_i,D_i)$ denote the outcome that would be observed given the treatment assignment $Z_i$ and the actual receipt of treatment $D_i$.
The notation we use implies the Stable Unit Treatment Value Assumption (SUTVA), which requires that the potential outcomes of the $i-$th participant are not related to the treatment assignment or receipt of the other participants \citep{Rubin1980}. A further assumption made in \citep{ImbensAngrist1994} is that treatment assignment is not related to the potential outcomes given the actual received treatment. They refer to this as the exclusion restriction assumption (ER). Under the ER, for every participant, there is only a pair of counterfactual outcomes, $Y_i(D_i)$, one for each value of the treatment $D_i$. We will thus denote $Y_i(0)$ the potential outcome that would be observed under the control treatment, $D_i=0$, and $Y_i(1)$ the potential outcome that would be observed under the active treatment, $D_i=1$. In reality, we can only observe one of the two outcomes, that is 
\begin{equation}\label{eq:Y}
    Y_i=D_iY_i(1)+(1-D_i)Y_i(0).     
\end{equation}
Given the SUTVA, we can define a causal effect of $Z$ on $Y$ as $Y_i(1,D_i(1))-Y_i(0,Z_i(0))$ and given the ER we can define a causal effect of $Z$ on $D$ as $D_i(1)-D_i(0)$ and a causal effect of $D$ on $Y$ as $Y_i(1)-Y_i(0)$, all at an individual level \citep{Angrist1996}. \citet{ImbensAngrist1994} make two additional assumptions. First, they assume that the average causal effect of $Z$ on $D$, $\mathbb{E}[D_i(1)-D_i(0)]$, is nonzero and secondly that $D_i(1)\geq D_i(0)$ for every participant, which they refer to as the monotonicity condition. The latter ensures that there are no participants who do the opposite of their assignment regardless of their assignment. They discuss how the monotonicity condition is fundamentally non-verifiable and needs to be considered in the context of a particular application \citep{Angrist1996}. An example of a design where the monotonicity condition is satisfied is the single-consent randomised encouragement design \citep{Zelen1979}. In this design, participants in the control group are prevented from receiving the active treatment and only participants assigned to the treatment are able to receive it.

By adapting this framework, participants are divided into 4 compliance types based on their potential treatment receipt values under both treatment groups.
Specifically, 
\begin{enumerate}
    \item If $D_i(1)=1$ and $D_i(0)=0$, the participant is a complier,
    \item If $D_i(1)=D_i(0)=0$, the participant is a never-taker,
    \item If $D_i(1)=D_i(0)=1$, the participant is an always-taker, 
    \item If $D_i(1)=0$ and $D_i(0)=1$, the participant is a defier.
\end{enumerate}
We summarise the compliance types based on the potential treatment receipt in Table \ref{tab:Table1}. The later three make up for the group of non-compliers in the study. Under the monotonicity assumption, defiers are excluded from the study, thus the non-compliers group consists of never-takers and always-takers. This is a conceptual division of the study participants because in reality we can only observe the actual receipt of treatment for each individual, $D_i$, which means that knowledge of the compliance status of the participants is incomplete \cite{LittleandYau1998}. 

In practice, we observe the following three quantities for each individual, the treatment assignment, the actual receipt of treatment and the observed outcome, i.e. the three dimensional vector $(Z_i,D_i,Y_i)$. Based on the observed quantities, in the treatment group, we cannot distinguish between compliers and always-takers and similarly, in the control group, we cannot distinguish between compliers and never-takers, assuming defiers are excluded. As a result, the observed non-compliers consist of never-takers from the treatment group and always-takers from the control group. The possible compliance types based on the observed values of $Z$ and $D$ excluding defiers are summarised in Table \ref{tab:Table2}. 

\begin{table}
    \centering
    \caption{Conceptual Compliance Types}
    \begin{tabular}{cc|cc}
    \hline
    \toprule
    & & \multicolumn{2}{c}{$D_i(0)$} \\ 
    & & 0 & 1 \\
    \midrule
    \multirow{2}{*}{$D_i(1)$} & 0 & Never-Taker & Defier \\
    & 1 & Complier & Always-Taker \\
    \bottomrule
    \hline
    \end{tabular}
    \label{tab:Table1}
\end{table}

Based on the conceptual division of participants,  \citet{ImbensAngrist1994} define the Local Average Treatment Effect (LATE) or Complier Average Causal Effect (CACE) as, 
\begin{equation}\label{eq:CACE}
\mathbb{E}[(Y_i(1)-Y_i(0))|D_i(1)-D_i(0)=1],
\end{equation}
or equivalently $\mathbb{E}[(Y_i(1)-Y_i(0))|C=1]$, where $C$ is a binary indicator of compliance, $C=1$ corresponds to the subgroup of compliers and $C=0$ to the subgroup of non-compliers. Under this notation, the Average Treatment Effect (ATE) across all participants is defined as, $\mathbb{E}[Y_i(1)-Y_i(0)]$.

\begin{table}[htbp]
    \centering
    \caption{Potential Observed Compliance Types Excluding Defiers}
    \begin{tabular}{cc|cc}
    \hline
    \toprule
    & & \multicolumn{2}{c}{$Z_i$} \\ 
    & & 0 & 1 \\
    \midrule
    \multirow{2}{*}{$D_i$} & 0 & Complier/Never-Taker & Never-Taker \\
    & 1 & Always-Taker & Complier/Always-Taker \\
    \bottomrule
    \hline
    \end{tabular}
    \label{tab:Table2}
\end{table}

This proposed framework is a special case of principal stratification and CACE is an example of a principal stratum estimand \citep{Frangakis2002Principal-Strat}. Principal stratification with respect to a posttreatment variable (in this case $D$) was originally formulated by \citet{Frangakis2002Principal-Strat} and consists of partition of the study participants into subsets such that all participants have the same vector of values $(D_i(0),D_i(1))$ within each subset. Each subset created is called a principal stratum and the comparison of potential outcomes under the control versus the active treatment within this stratum defines a principal effect.  \citet{Frangakis2002Principal-Strat} proved that any principal effect is a well-defined causal effect, since it always compares potential outcomes under two treatments on a common set of units. Excluding defiers, the study population can be stratified into three principal strata, the compliers, the never-taker and the always-takers. Under the exclusion restriction, the principal effect for never-takers and always-takers is zero since these participants always receive the same treatment regardless of their allocated one \citep{LittleandYau1998}.

\subsection{Per-Protocol Analysis}
One of the simplest analyses employed in practice for estimation of CACE is the Per-Protocol Analysis. A Per-Protocol analysis excludes participants who did not comply with their original treatment allocation and considers only participants who strictly complied with the study protocol. The population included in a Per-Protocol Analysis is sometimes referred to as the Per-Protocol subpopulation \citep{Tripepietal.2020}. In essence, the Per-Protocol subpopulation consists of the observed compliers as defined in Section \ref{subsec:CACE}.   

Observed compliers in the treatment group ($Z=1$) receive the active treatment ($D=1$) and observed compliers in the control group ($Z=0$) receive the control treatment ($D=0$). Participants with $Z_i\neq D_i$ are the observed non-compliers (never-takers if $Z=1$ and always-takers if $Z=0$). The Per-Protocol estimator is defined as the difference in outcome averages among observed compliers in the treatment group versus observed compliers in the control group, i.e. 
\begin{equation}\label{eq:PPE}
  \hat{\text{PPE}}=\bar{Y}_{Z=D=1}-\bar{Y}_{Z=D=0},    
\end{equation}
where $\bar{Y}_{Z=D=1}$ and $\bar{Y}_{Z=D=0}$ correspond to the observed mean of $Y$ among compliers in the treatment and the control group respectively.

The major disadvantage of the Per-Protocol method is that the Per-Protocol subpopulation does not preserve the benefits of randomisation as participants in the two treatment groups of the Per-Protocol subpopulation may no longer be comparable for known or unknown covariates. As a consequence, the Per-Protocol analysis usually yields biased estimates in RCTs regardless of whether it is estimating the ATE or CACE \citep{Kamper2021}. 

\subsection{Instrumental Variables Analysis}\label{IV theory}
The original method proposed in \citep{ImbensAngrist1994} and extended in detail in \citep{Angrist1996} to estimate CACE is the Instrumental Variables method. Specifically, if the assumptions discussed in Section \ref{subsec:CACE} are satisfied (Random Assignment, SUTVA, ER, Nonzero Average Causal Effect of $Z$ on $D$ and Monotonicity), then the treatment assignment variable, $Z$, is called an Instrument or an Instrumental Variable (IV). Provided the assumptions hold, the IV estimand is defined as, \begin{equation}\label{eq:IV0}
    \text{IV}=\frac{\mathbb{E}[Y_i|Z_i=1]-\mathbb{E}[Y_i|Z_i=0]}{\mathbb{E}[D_i|Z_i=1]-\mathbb{E}[D_i|Z_i=0]},
\end{equation} and the IV estimator as, 
\begin{equation}\label{eq:IV1}
    \hat{\text{IV}}=\frac{\bar{Y}_{Z=1}-\bar{Y}_{Z=0}}{\bar{D}_{Z=1}-\bar{D}_{Z=0}},  
\end{equation}
where $\bar{Y}_{Z}$ and $\bar{D}_{Z}$ correspond to the observed mean of $Y$ and $D$ for treatment assignment $Z$ respectively. In (\ref{eq:IV0}) becomes apparent the assumption of non-zero effect of $Z$ on $D$ discussed in Section \ref{subsec:CACE}. 

Imbens and Angrist \citep{ImbensAngrist1994} proved that the IV estimand coincides with CACE, i.e. $$\mathbb{E}[(Y_i(1)-Y_i(0))|C=1]=\frac{\mathbb{E}[Y_i|Z_i=1]-\mathbb{E}[Y_i|Z_i=0]}{\mathbb{E}[D_i|Z_i=1]-\mathbb{E}[D_i|Z_i=0]}.$$

The IV estimator is also referred to as the two stage least squares (2SLS) estimator because it can be obtained using a two-stage least squares procedure. First, we regress the treatment receipt variable $D_i$ on the instrument $Z_i$ to obtain the following estimated regression function, $$\hat{D_i}=\hat{\alpha}_0 + \hat{\alpha}_1Z_i.$$ Then, we regress the observed outcome $Y_i$ on the predicted treatment receipt value, $\hat{D_i}$, 
\begin{equation}\label{eq:Stage2}
    Y_i=\hat{\beta}_0+\hat{\beta}_1\hat{D_i}.
\end{equation}
The IV estimator is equal to $\hat{\beta}_1$. The same result can be obtain by using the following two regression models,
$$Y_i=\alpha_0+\alpha_1 Z_i+\epsilon_i \; \; \text{and} \; \; D_i=\beta_0+\beta_1Z_i+v_i.$$ Assuming that $Z_i$ is independent of $\epsilon_i$ (ER Assumption) and $v_i$ and that the covariance between the treatment receipt $D_i$ and assignment $Z_i$ is nonzero, i.e. $\text{cov}(D_i,Z_i)\neq0$, the IV estimator is equal to the ratio, 
\begin{equation}\label{eq:IV2}
    \hat{\text{IV}}=\frac{\hat{\alpha}_1}{\hat{\beta}_1},  
\end{equation}
where $\hat{\alpha}_1$ and $\hat{\beta}_1$ are the least squares estimators of $\alpha_1$ and $\beta_1$ respectively. 

Both, in (\ref{eq:IV1}) and (\ref{eq:IV2}), the denominator is an estimator of the proportion of true compliers \citep{Angrist1996} and the numerator is an estimator of the effect of treatment assignment $Z$ on the outcome $Y$, which is usually called the Intention-to-Treat (ITT) effect and is defined as, $$\mathbb{E}[Y_i|Z_i=1]-\mathbb{E}[Y_i|Z_i=0].$$

\subsection{The CPAP study}
The simulations we employ were motivated by an RCT that compared the CPAP treatment with usual antenatal care for women with high-risk pregnancy and obstructive sleep apnea (OSA) \citep{Tantrakuletal.2023}. One of the two primary endpoints was the modulation of diastolic blood pressure (DBP). The CPAP treatment is administered through a machine which delivers continuous air through the mouth and/or the nose to help keep an individual's airways open during sleep. In the study conducted by \citet{Tantrakuletal.2023}, the observed compliance rate in the CPAP treatment group was approximately $33\%$. 340 participants were needed to detect a lowering effect of $2.5$ mmHg with $80\%$ power and $5\%$ significance level. Only 310 participants were considered in the final study and were randomly assigned to receive either the CPAP treatment or usual care (153 in the treatment group and 157 in the control group). An ITT analysis showed a $2.2$ decrease in DBP for participants in the CPAP group, while a Per-Protocol Analysis showed a $3.2$ decrease in DBP for compliers in the CPAP group. The standard deviation of the Per-Protocol effect was estimated to be $1.24$. Based on this estimate, at least $227$ simulations are needed to produce an effect estimate of $5\%$ accuracy using the standard formula from \citep{Burtonetal.2006}. The derivation of this can be found in the Supplementary Material. We decide to generate $500$ simulations for each trial we consider.

\subsection{Simulation Structure}
On the basis of the CPAP study, we simulate hypothetical RCTs where participants are randomly assigned to the CPAP treatment or usual care by a 1:1 allocation ratio. In these simulations we chose to consider 350 participants (175 at each arm). Let us denote $Z$ the assignment variable which was generated as $Z\sim \text{Bernoulli}(0.5)$ for every participant. The parameters for generating hypothetical participants were estimated from the original CPAP study \citep{Tantrakuletal.2023}. Specifically, the mean DBP for participants in the usual care group at baseline was $73.9$ mmHg with $0.7$ standard deviation. We thus generate $Y(0)$ from a Normal distribution with mean $74$ and standard deviation $1$, i.e. $Y(0)\sim N(74,1)$. For the treatment effect in our primary simulation analysis, we consider a decrease of $5$ mmHg in DBP over the usual care, i.e. the treatment effect is defined as $\tau=\mu_1-\mu_0=-5$, where $\mu_1$, $\mu_0$ the mean of the potential outcomes under treatment and control respectively. This corresponds to the ATE across all individuals. In summary, we generate for each individual the following three quantities, $Y(0)\sim N(74,1)$, $Y(1)\sim N(69,1)$ and $Z\sim\text{Bernoulli}(0.5)$.

\begin{figure*}
    \centering
    \includegraphics[width=0.5\linewidth]{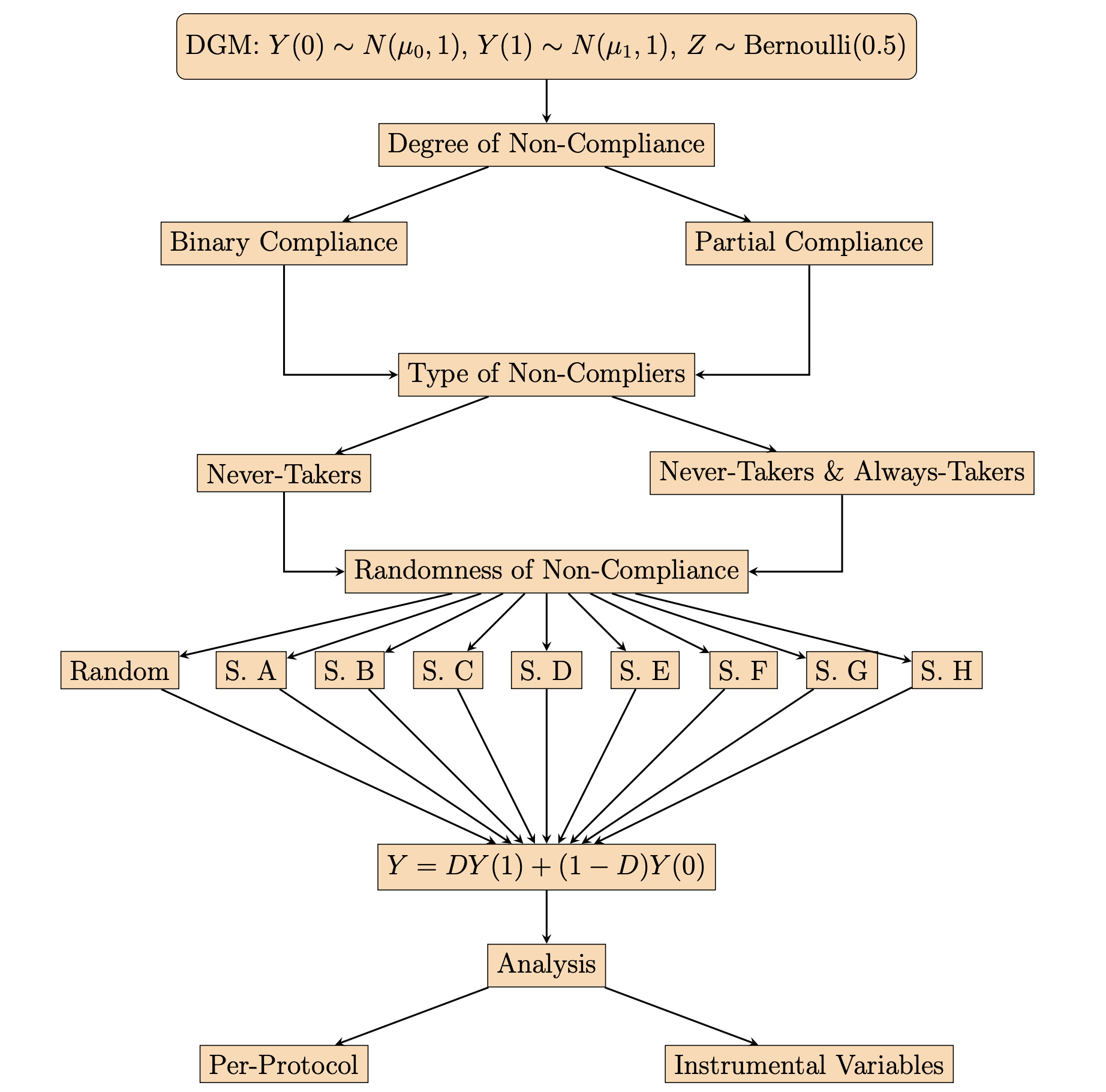}
    \caption{\textbf{Flowchart of the simulation procedure.} DGM: Data Generating Mechanism, S.: Scenario, $Y(0):$ The counterfactual outcome for a participant in the control group, $Y(1):$ The counterfactual outcome for a participant in the treatment group, $Z:$ Treatment assignment indicator, $D:$ Degree of Compliance with the Treatment.}
    \label{fig:flowchart}
\end{figure*}

We simulate different non-compliant scenarios by varying the following factors: ($1$) the degree of non-compliance, ($2$) the type of non-compliers and ($3$) the randomness of non-compliance. Our simulation framework is shown in Figure \ref{fig:flowchart}. Before we describe the three varying factors in detail we make a final assumption. Since the CPAP treatment is administered during sleep and the average sleep time for pregnant women is 7-9 hours, we assume that the complete use of the CPAP treatment ($100\%$) is defined to be 8 hours. 

\paragraph{Degree of Non-Compliance}
Based on the previous assumption, we vary the first non-compliant factor, the degree of non-compliance, i.e. the proportion of interventional components a participant did not receive according to the protocol or in this case the number of hours a participant did not use the CPAP treatment. Specifically, we consider two cases, one where compliance is binary, i.e. participants can either receive $100\%$ of the CPAP treatment (8 hours) or none of it (0 hours), and the second where compliance is partial, i.e. participants can receive some parts of the CPAP treatment, e.g. receive CPAP for 4 hours. For the partial compliance case, we consider only multiples of 2, i.e. a participant can use the CPAP treatment for 0, 2, 4, 6 or 8 hours. Let us denote $D$ the proportion of CPAP use. For the binary compliance case, $D$ can either take the value $1$, which corresponds to 8 hours ($100\%$) of CPAP use or $0$ which corresponds to 0 hours ($0\%$) of CPAP use. For the partial compliance case, $D$ can take the values $1$, $3/4$, $1/2$, $1/4$ or $0$ for 8, 6, 4, 2 and 0 hours of CPAP use respectively. In the main simulation analysis, compliers are defined as the participants who received the complete CPAP treatment. i.e. 8 hours.

\paragraph{Type of Non-Compliers} The second factor we vary is the type of non-compliers. Again, we consider two cases, one where non-compliers are only never-takers and one where non-compliers can either be never-takers or always-takers. The former mimics the scenario where non-compliance can occur only in the treatment group, i.e. participants who were assigned to the CPAP treatment ($Z=1$) ended up not receiving it ($D=0$) and participants in the control group ($Z=0$) had no other choice but to comply with their assigned treatment ($D=0$), while the latter mimics the scenario where non-compliance can occur in either treatment group. For the binary compliance case, we simulate never-takers and always-takers following the conceptual framework of \cite{ImbensAngrist1994}, based on the participants' potential receipt of treatment under their allocated treatment, as described in Table \ref{tab:Table1}. For the partial compliance case, this conceptual framework cannot be applied and instead we simulate never-takers and always-takers as we would actually observe them in practice, i.e. never-takers can only be observed in the treatment group and always-takers can only be observed in the control group.

\paragraph{Randomness of Non-Compliance} Non-compliant behavior could be random or dependent on participants' conditions (non-random). For the former, we randomly select a subset of participants to be non-compliers. For the binary compliance case, we consider $40\%$ of the study population to be non-compliers. In the presence of both never-takers and always-takers, we consider $25\%$ to be never-takers ($D(1)=D(0)=0$) and $15\%$ to be always-takers ($D(1)=D(0)=1$). For the partial compliance case we assume the following, 
\begin{itemize}
    \item For the control group: $50\%$ comply with their allocated treatment ($D=0$), $20\%$ take CPAP for 2 hours ($D=1/4$), $10\%$ take CPAP for 4 hours ($D=1/2$), $10\%$ take CPAP for 6 hours ($D=3/4$) and $10\%$ take the complete CPAP treatment ($D=1$),
    \item For the treatment group: $50\%$ comply with their allocated treatment ($D=1$), $20\%$ take CPAP for 6 hours ($D=3/4$), $10\%$ take CPAP for 4 hours ($D=1/2$), $10\%$ take CPAP for 2 hours ($D=3/4$) and $10\%$ do not take CPAP at all ($D=0$).
\end{itemize}

When non-compliance depends on the participants' conditions, we consider eight different non-random scenarios. Specifically, the scenarios we consider are,
\begin{itemize}
    \item \textbf{Scenario A:} Participants with good conditions would never receive the treatment,
    \item \textbf{Scenario B:} Participants with bad conditions would never receive the treatment,
    \item \textbf{Scenario C:} Participants with good conditions would always receive the treatment,
    \item \textbf{Scenario D:} Participants with bad conditions would always receive the treatment,
    \item \textbf{Scenario E:} Participants with bad conditions would always receive the treatment, while participants with good conditions would never receive the treatment,
    \item \textbf{Scenario F:} Participants with bad conditions would never receive the treatment, while participants with good conditions would always receive the treatment,
    \item \textbf{Scenario G:} Participants with good conditions would never receive the treatment, while participants with severe bad conditions would always receive the treatment. Additionally, participants with moderately bad conditions would always receive three quarters of the treatment, participants with moderate to mild bad conditions would always receive half of the treatment and participants with mild bad conditions would always receive a quarter of the treatment,
    \item \textbf{Scenario H:} Participants with bad conditions would never receive the treatment, while participants with extremely good conditions would always receive the treatment. Additionally, participants with moderately good conditions would always receive three quarters of the treatment, participants with moderate to mild good conditions would always receive half of the treatment and participants with mild good conditions would always receive a quarter of the treatment.
\end{itemize}

We provide a formal definition of participants' conditions in the next paragraph. The first six scenarios (A-F) were studied by \citet{McNamee2009} and examined in detail by \citet{Yeetal.2014} in their simulations to estimate ATE under non-compliance, but they have not been examined in estimating CACE. We propose and examine two additional scenarios (G-H) that can only be applied in the partial compliance case. Scenarios C, D, E and F can be applied only in the presence of both never-takers and always-takers. For scenarios G and H, in the presence of never-takers only, the conditions are applied only to participants in the control group. Participants' conditions are considered to be associated with their potential outcomes under usual care, $Y(0)$ \citep{Yeetal.2014}. 

For scenarios A-F, participants are said to have good conditions if they satisfy the following condition, $Y(0) < \mu_0 - \delta$ and are said to have bad conditions if they satisfy $Y(0) > \mu_0 + \delta$, where $\mu_0$ is the mean of $Y(0)$ and $\delta$ is a parameter that indicates the degree of dependence on $Y(0)$. For scenario G we define severe bad conditions as $Y(0) > \mu_0 + \delta$, moderately bad conditions as $\mu_0 + \delta - 0.2 < Y(0) < \mu_0 + \delta$, moderate to mild bad conditions as $\mu_0 + \delta - 0.35 < Y(0) < \mu_0 + \delta - 0.2$ and mild bad conditions as $\mu_0 < Y(0) < \mu_0 + \delta - 0.35$. For scenario H we define extremely good conditions as $Y(0) < \mu_0 - \delta$, moderately good conditions as $\mu_0 - \delta < Y(0) < \mu_0 - \delta + 0.2$, moderate to mild good conditions as $\mu_0 - \delta + 0.2 < Y(0) < \mu_0 - \delta + 0.35$ and mild good conditions as $\mu_0 - \delta + 0.35 < Y(0) < \mu_0$. In our main simulation analysis, we consider $\delta=0.5$ for severe dependence on $Y(0)$ and we perform a sensitivity analysis using $\delta=1$ for mild dependence on $Y(0)$. For scenarios G and H, when $\delta=1$, for the lower and upper bounds of the mild conditions respectively, which are both equal to $\mu_0$, we add $0.5$.

\paragraph{Simulation Analysis}
So far we have introduced the data generating mechanism and the varying non-compliant factors that we use in our simulations. We will now describe the complete process of our simulation analysis by combining all the above. We begin with the binary compliance case. First, we generate the potential outcomes $Y(0)$, $Y(1)$ and randomly assign participants ($Z$) to either treatment or control. Under the random non-compliance scenario, we let a randomly selected subset of $40\%$ of the participants to be non-compliers. This corresponds to the random non-compliance scenario. If the simulation allows only for never-takers, the $40\%$ of non-compliers subgroup consists only of never-takers ($D(1)=D(0)=0$). If the simulation allows for both never-takers and always-takers, the $40\%$ of non-compliers subgroup is divided into $25\%$ never-takers ($D(1)=D(0)=0$) and $15\%$ always-takers ($D(1)=D(0)=1$). For the non-random non-compliant scenarios, we apply, additionally to the randomly selected $40\%$ non-complier subset, one of the eight scenarios (A-H). This leads to an increased number of non-compliers. Then, we calculate the observed values of $D$ and $Y$ using (\ref{eq:D}) and (\ref{eq:Y}) and finally, we perform the Per-Protocol and IV analyses.

The simulation analysis for the partial compliance case is very similar with the only difference being in the random non-compliance scenario since $D$ is not binary in this case. First, we start with the data generating mechanism ($Y(0)$, $Y(1)$, $Z$). Then, we randomly assign participants to their proportions of CPAP receipt using the specified proportions previously descirbed. In the presence of never-takers only, we assign $D=0$ to every participant in the control group. Under the non-random non-compliant scenarios, we apply additionally, the conditions described in each scenario. The remaining steps are the same as in the binary compliance case. 

The results from each simulation run were derived using the Monte Carlo estimates of the 500 simulations. The Per-Protocol and the IV methods were compared by the bias, mean squared error (MSE) and $95\%$ coverage \citep{Burtonetal.2006}. 

\paragraph{Sensitivity Analyses}
We conduct a number of sensitivity analyses in order to consider a wider range of different conditions and scenarios apart from our main simulation. For both binary and partial compliance cases, we consider a sensitivity analysis on the severity of the non-random non-compliant scenarios by applying the mild dependent conditions. Also, a sensitivity analysis is conducted on the treatment effect which is now set equal to $0$. For the partial compliance scenario we consider different thresholds of compliance, i.e. participants are considered to be compliers if they have complied with at least a specified number of hours of CPAP. We employ the four following cases,  

\begin{itemize}
 \item CASE I: Compliers are defined as participants who take the whole CPAP treatment (8 hours),
 \item CASE II: Compliers are defined as participants who take at least $75\%$ of the CPAP treatment (at least 6 hours),
 \item CASE III: Compliers are defined as participants who take at least half ($50\%$) of the CPAP treatment (at least 4 hours), and
 \item CASE IV: Compliers are defined as participants who take any part of the CPAP treatment (more than 0 hours)
\end{itemize}

Case I is applied on the main simulation analysis and we consider sensitivity analyses on the other three (II-IV). Finally, we consider sensitivity analyses on the proportions of non-compliers. For the binary compliance case, we consider a larger proportion of non-compliers ($60\%$). In the presence of both never-takers and always-takers we consider $40\%$ as never-takers and $20\%$ as always-takers.

For the partial compliance case, we consider the following proportions. For the control group, $60\%$ comply with the control treatment ($D = 0$), $10\%$ take CPAP for 2 hours ($D = 1/4$), $10\%$ take CPAP for 4 hours (D = 1/2), $10\%$ take CPAP for 6 hours ($D = 3/4$) and $10\%$ take CPAP for 8 hours ($D = 1$). For the treatment group, $40\%$ comply with the allocated treatment ($D = 1$), $10\%$ take CPAP for 6 hours ($D = 3/4$), $10\%$ take CPAP for 4 hours
($D = 1/2$), $10\%$ take CPAP for 2 hours ($D = 1/4$) and $30\%$ do not take CPAP at all ($D = 0$).


\section{Results}\label{sec:results}
In Tables \ref{tab:res} and \ref{tab:resul} we summarise the results from our main simulation analysis in the presence of both never-takers and always-takers and of never-takers only respectively.

\subsection{Results for the Binary Compliance case}
We discuss first the results from the binary compliance case. When non-compliance is random, CACE coincides with the ATE and both the Per-Protocol and the IV estimators are unbiased for both types of non-compliers. When non-compliance is non-random and depends on the potential outcomes under usual care, the IV estimator is unbiased for CACE for all non-random Scenarios imposed (A-F) regardless of the type of non-compliers. For Scenarios E and F, in the presence of never-takers and always-takers, CACE coincides with the ATE and hence the IV estimator is also unbiased for the ATE. The Per-Protocol estimator is unbiased for CACE only for the non-random Scenarios C and D in the presence of never-takers and always-takers. Specifically, the Per-Protocol estimator is unbiased for CACE when participants with good or bad conditions would always receive the treatment. In the presence of never-takers only, the Per-Protocol estimator is biased for CACE for the two non-random scenarios (A and B), but it is unbiased for the ATE in both cases. In general, the IV estimator provides lower MSE values and higher $95\%$ coverage results than the Per-Protocol one for all non-random scenarios.

The sensitivity analyses on the treatment effect and the proportion of non-compliers did not show any significant difference to the results from the main simulation apart from one case. Specifically, for the sensitivity analysis on the proportion of non-compliers, for Scenarios E and F the proportions of true non-compliers were high ($85\%$ true non-compliers) which resulted in the IV estimator being slightly biased for the two cases ($0.13$ and $0.06$ bias respectively). Finally, for the sensitivity analysis where we considered the mild non-random non-compliance scenarios ($\delta=1$), the Per-Protocol estimator was less biased than in the severe case ($\delta=0.5$) for all scenarios and both types of non-compliers, while the IV estimator remained unbiased regardless of any factor related to compliance.

\subsection{Results for the Partial Compliance case}
We continue with the results for the partial compliance case. When non-compliance is random, CACE coincides with the ATE and both the Per-Protocol and the IV estimators are unbiased regardless of the type of non-compliers. When non-compliance is non-random, both the Per-Protocol and the IV estimators are biased for CACE for both types of non-compliers. In the presence of never-takers and always-takers and for the non-random Scenarios E, F, G and H, the IV estimator is significantly less biased than the Per-Protocol estimator. Particularly, for Scenario F, the IV estimator is approximately unbiased for CACE (bias equal to $-0.04$). On the other hand, the Per-Protocol estimator is significantly less biased than the IV estimator for Scenarios A and B, where participants with good or bad conditions would never receive the active treatment. For the two remaining scenarios (C and D), where participants with good or bad conditions would always receive the treatment, both estimators are similar in terms of bias. For Scenarios A-F, CACE coincides with the ATE and the results presented for CACE can be generalised for the ATE as well. For Scenarios G and H, CACE differs from the ATE but both estimators are also biased for the ATE. For all eight Scenarios, the MSE values of the Per-Protocol estimates are smaller than the MSE values of the IV estimates, but the IV estimator provides better $95\%$ coverage results in general. In the presence of never-takers only, for all four non-random Scenarios, the Per-Protocol estimator is less biased for CACE than the IV estimator and provides lower MSE values along with higher $95\%$ coverage. Finally, we observe that the Per-Protocol estimator is always unbiased for the ATE in the presence of never-takers.

For Scenarios G and H when always-takers and never-takers were allowed, we observe from Table \ref{tab:res}, that the within simulation standard error for the bias and the MSE values of the IV estimator are way higher than any other simulation. This is by definition of the two scenarios. Specifically, the conditions we impose for participants to receive or not the active treatment or parts of it, in a small number of simulation runs, completely disregards the original treatment allocation which results in violation of the condition of non-zero covariance between the instrument and the treatment receipt ($D$) in the 2SLS regression described in Section \ref{IV theory}. As a result, estimates in the second stage of the 2SLS (\ref{eq:Stage2}) are "divergent"\footnote{The term divergent refers to estimates that have very high bias relative to CACE.} and have high standard errors. Specifically, for Scenario G, we identified 13 "divergent" results out of 500 simulations ($2.6\%$) and for Scenario H, we identified 12 ($2.4\%$). This prompted us to perform a secondary analysis where we excluded these values from the Monte Carlo estimation of the IV estimator. For Scenario G, the bias of the IV estimator excluding the "divergent" results was $-0.22$ with an estimated within simulation standard error equal to $0.085$ and MSE equal to $3.5$. For Scenario H, the IV estimator was unbiased with an estimated within simulation standard error equal to $0.09$ and MSE equal to $4$. Both the standard error and the MSE values are substantially lower when excluding "divergent" results. Whether excluding the "divergent" results or not, the IV estimator is, in both cases, less biased than the Per-Protocol estimator for Scenarios G and H.

The sensitivity analysis on the treatment effect and the proportion of non-compliers did not show any significant difference from the results of the main simulation. A notable observation we make is that for the sensitivity analysis on the proportion of non-compliers, the MSE values of the IV estimates were substantially higher, especially in the presence of both never-takers and always-takers. When considering the mild non-random non-compliant scenarios, all estimates were substantially less biased for both the ATE and CACE. Specifically, in the presence of never-takers and always-takers, the IV estimator was unbiased for CACE for Scenarios E, F, G and approximately unbiased for Scenario H. Regarding the sensitivity analysis on the different compliance thresholds, we observe that the IV estimator does not change with the thresholds, since the 2SLS regression considers all participants in the study and is not restricted to the participants defined by the threshold. Additionally, when non-compliance is random, the IV estimator is unbiased for CACE regardless of the threshold. In the presence of always-takers and never-takers, the Per-Protocol estimator becomes more biased as we include participants who took lower doses of the CPAP treatment as compliers except for Scenarios F and H, where it becomes less biased with each threshold. The Per-Protocol estimator was unbiased for CACE in 3 cases, $(1)$ Scenario B, Threshold Case II, $(2)$ Scenario C, Threshold Case III and $(3)$ Scenario H, Threshold Case IV. To decide whether these results are meaningful or simply occur by chance we run a secondary analysis (results not provided) on the compliance thresholds to the sensitivity analyses on the different proportions. The secondary analysis showed inconsistency between the previous results as the Per-Protocol estimator was biased for all three cases. Consequently, we conclude that the results from $(1)-(3)$ occur simply by chance and the Per-Protocol estimator is biased for all threshold cases when non-compliance is non-random.
In the presence of never-takers only, the IV estimator is unbiased for CACE regardless of the compliance threshold when non-compliance is random and it is generally less biased than the Per-Protocol for all thresholds.

  \begin{table*}
  \centering
  \renewcommand{\arraystretch}{0.7}
  \caption {\label{tab:res} Results when Never-Takers and Always-Takers were allowed (Treatment Effect $= -5$).} 
  \label{tab:results}
  \begin{adjustbox}{max width=\textwidth}
  \begin{tabular}{cccccc|cccccc}
  \toprule
  \multicolumn{6}{c}{Binary Compliance} & \multicolumn{6}{c}{Partial Compliance} \\
  \cmidrule(lr){1-6} \cmidrule(lr){7-12}
  Scenario & Method & Estimate & Bias (SE) & MSE & Coverage(\%) & Scenario & Method & Estimate & Bias (SE) & MSE & Coverage(\%) \\
  \midrule
  Random & CACE & -5.01 & - & - & - & Random & CACE & -4.99 & - & - & - \\
  & PP & -5 & -0.01 (0.004) &0.008 & 98\% & & PP & -5 & 0.01 (0.005) & 0.013 & 97\% \\
  & IV & -5.01 & 0 (0.007) & 0.022 & 98\% & & IV & -5.01 & 0.02 (0.009) & 0.037 & 97\% \\
  \midrule
  Scenario A & CACE & -5.51 & - & - & - & Scenario A & CACE & -5 & - & - & - \\
  & PP & -4.95 & -0.56 (0.005) & 0.333 & 0\% & & PP & -4.73 & -0.27 (0.005) & 0.083 & 68\% \\
  & IV & -5.52 & 0.01 (0.012) & 0.07 & 97\% & & IV & -5.55 & 0.55 (0.015) & 0.426 & 76\% \\
  \midrule
  Scenario B & CACE & -4.49 & - & - & - & Scenario B & CACE & -4.99 & - & - & - \\
  & PP & -5.06 & 0.57 (0.005) & 0.329 & 0\% & & PP & -5.27 & 0.28 (0.005) & 0.092 & 62\% \\
  & IV & -4.48 & -0.01 (0.012) & 0.069 & 97\% & & IV & -4.48 & -0.51 (0.015) & 0.374 & 81\% \\
  \midrule
  Scenario C & CACE & -5.51 & - & - & - & Scenario C & CACE & -5 & - & - & - \\
  & PP & -5.51 & 0 (0.005) & 0.011 & 96\% & & PP & -5.51 & 0.51 (0.005) & 0.279 & 1\% \\
  & IV & -5.51 & 0 (0.009) & 0.043 & 98\% & & IV & -5.53 & 0.53 (0.013) & 0.364 & 57\% \\
  \midrule
  Scenario D & CACE & -4.49 & - & - & - & Scenario D & CACE & -4.99 & - & - & - \\
  & PP & -4.49 & 0 (0.004) & 0.009 & 97\% & & PP & -4.49 & -0.5 (0.005) & 0.261 & 0\% \\
  & IV & -4.49 & 0 (0.009) & 0.043 & 97\% & & IV & -4.5 & -0.49 (0.012) & 0.316 & 64\% \\
  \midrule
  Scenario E & CACE & -5 & - & - & - & Scenario E & CACE & -4.99 & - & - & - \\
  & PP & -4.45 & -0.55 (0.006) & 0.324 & 0\% & & PP & -4.3 & -0.69 (0.005) & 0.499 & 0\% \\
  & IV & -5.03 & 0.03 (0.021) & 0.214 & 97\% & & IV & -5.12 & 0.13 (0.029) & 0.44 & 97\% \\
  \midrule
  Scenario F & CACE & -5 & - & - & - & Scenario F & CACE & -4.99 & - & - & - \\
  & PP & -5.56 & 0.56 (0.006) & 0.325 & 0\% & & PP & -5.71 & 0.72 (0.005) & 0.521 & 0\% \\
  & IV & -4.95 & -0.05 (0.02) & 0.211 & 98\% & & IV & -4.95 & -0.04 (0.033) & 0.531 & 96\% \\
  \midrule
  Scenario G & CACE & - & - & - & - & Scenario G & CACE & -4.93 & - & - & - \\
  & PP & - & - & - & - & & PP & -4.07 & -0.86 (0.006) & 0.767 & 0\% \\
  & IV & - & - & - & - & & IV & -5.16 & 0.23 (0.207) & 21 & 96\% \\
  \midrule
  Scenario H & CACE & - & - & - & - &Scenario H & CACE & -5.05 & - & - & - \\
  & PP & - & - & - & - & & PP & -5.93 & 0.88 (0.006) & 0.792 & 0\% \\
  & IV & - & - & - & - & & IV & -5.23 & 0.18 (0.262) & 34 & 95\% \\
  \bottomrule
  \end{tabular}
  \end{adjustbox}
  \begin{tabbing}
  CACE: True Effect in Compliers, PP: Per-Protocol, IV: Instrumental Variables, \\
  SE: Within simulation standard error of bias, MSE: Mean Squared Error
  \end{tabbing}
  \end{table*}

\begin{table*}
\centering
\renewcommand{\arraystretch}{0.7}
\caption{\label{tab:resul}Results when only Never-Takers were allowed (Treatment Effect is $-5$).}
\begin{adjustbox}{max width=\textwidth}
\begin{tabular}{cccccc|cccccc}
\toprule
\multicolumn{6}{c}{Binary Compliance} & \multicolumn{6}{c}{Partial Compliance} \\
\cmidrule(lr){1-6} \cmidrule(lr){7-12}
Scenario & Method & Estimate & Bias (SE) & MSE & Coverage(\%) & Scenario & Method & Estimate & Bias (SE) & MSE & Coverage(\%) \\
\midrule
Random & CACE & -5.01 & - & - & - & Random & CACE & -5 & - & - & - \\
& PP & -5 & -0.01 (0.004) & 0.008 & 99\% & & PP & -4.99 & -0.01 (0.004) & 0.009 & 99\% \\
& IV & -5 & -0.01 (0.007) & 0.021 & 97\% & & IV & -4.99 & -0.01 (0.005) & 0.011 & 99\% \\
\midrule
Scenario A & CACE & -5.51 & - & - & - & Scenario A & CACE & -5.13 & - & - & - \\
& PP & -5 & -0.51 (0.005) & 0.273 & 1\% & & PP & -4.99 & -0.14 (0.005) & 0.032 & 90\% \\
& IV & -5.51 & 0 (0.011) & 0.059 & 97\% & & IV & -5.52 & 0.39 (0.008) & 0.186 & 64\% \\
\midrule
Scenario B & CACE & -4.49 & - & - & - & Scenario B & CACE & -4.87 & - & - & - \\
& PP & -5 & 0.51 (0.005) & 0.267 & 2\% & & PP & -4.99 & 0.12 (0.005) & 0.029 & 91\% \\
& IV & -4.49 & 0 (0.011) & 0.064 & 97\% & & IV & -4.48 & -0.39 (0.009) & 0.184 & 66\% \\
\midrule
Scenario G & CACE & - & - & - & - & Scenario G & CACE & -5.23 & - & - & - \\
& PP & - & - & - & - & & PP & -4.99 & -0.24 (0.005) & 0.069 & 64\% \\
& IV & - & - & - & - & & IV & -5.64 & 0.41 (0.008) & 0.202 & 49\% \\
\midrule
Scenario H & CACE & - & - & - &- & Scenario H & CACE & -4.76 & - & - & - \\
& PP & - & - & - & - & & PP & -4.99 & 0.23 (0.005) & 0.065 & 66\% \\
& IV & - & - & - & - & & IV & -4.36 & -0.4 (0.008) & 0.197 & 52\% \\
\bottomrule
\end{tabular}
\end{adjustbox}
\begin{tabbing}
CACE: True Effect in Compliers, PP: Per-Protocol, IV: Instrumental Variables, \\
SE: Within simulation standard error of bias, MSE: Mean Squared Error
\end{tabbing}
\end{table*}

\begin{figure*}[h]
    \centering
    \includegraphics[width=0.75\linewidth]{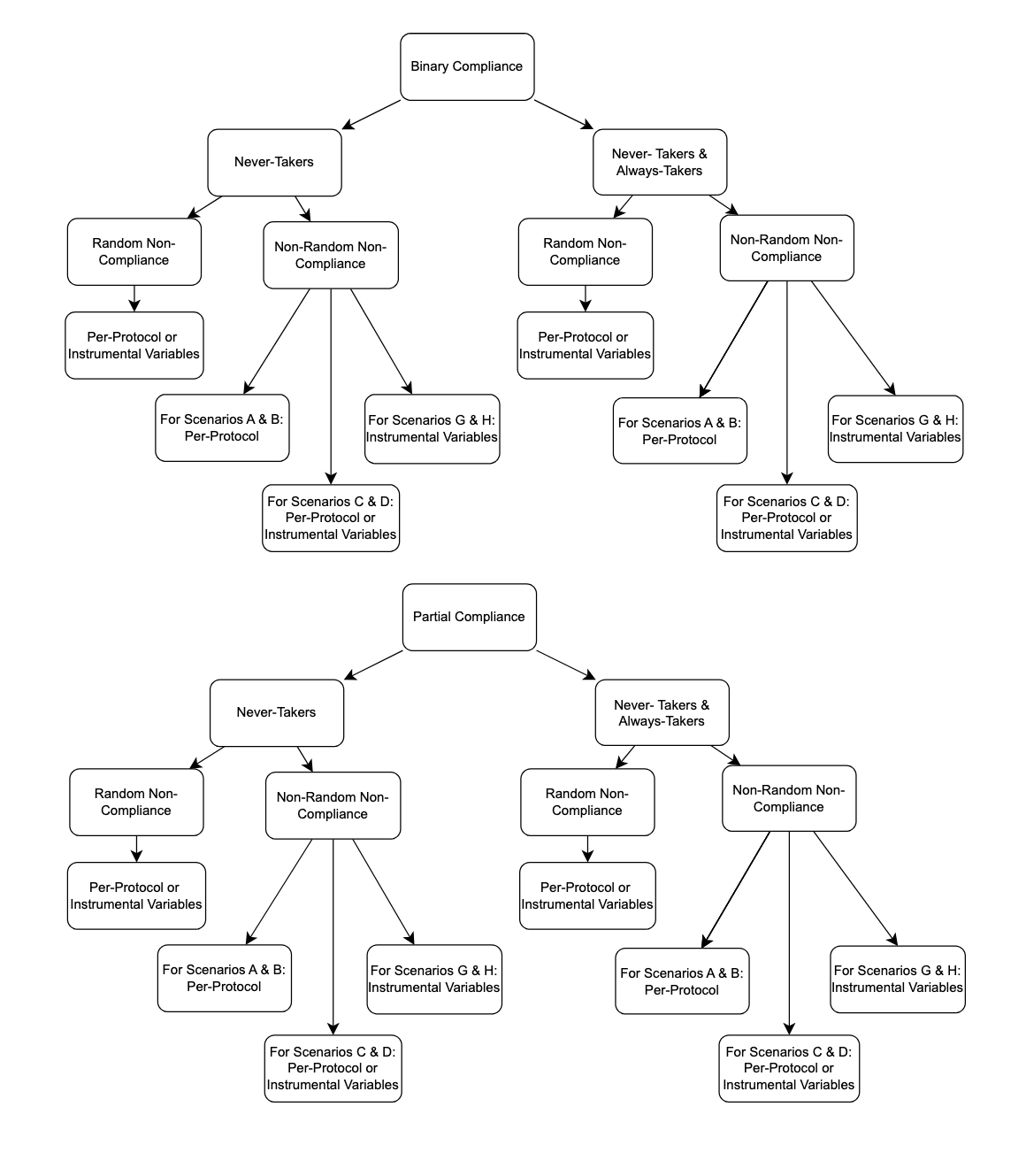}
    \caption{\textbf{A guide on the optimal method for estimating CACE for the various non-compliant scenarios.}
    On top, the optimal choice for the non-compliant scenarios when compliance is binary.
    Below, the optimal choice for the non-compliant scenarios when compliance is partial.}
    \label{fig:Choice}
\end{figure*}

In Figure \ref{fig:Choice} we provide a guide on which analysis is optimal or preferable to produce an unbiased or less biased estimate under the various scenarios we consider.

\section{Discussion}\label{sec:discussion}
In this paper, we have discussed the causal effect in the subgroup of compliers (CACE), which is of particular interest to clinical researchers for RCTs when non-compliance is present. We presented the CACE framework originally formulated in \citep{ImbensAngrist1994} along with the two most common methods employed in practice to estimate CACE, the Per-Protocol and the IV analyses. Our objective was to compare these two methods under various non-compliant scenarios usually encountered in RCTs. For this reason, we simulated hypothetical trials from a study on the CPAP treatment for women with high risk pregnancy and OSA by varying non-compliant factors such as the degree of non-compliance, the type of non-compliers and the randomness of non-compliance.

Our results showed that when compliance is binary, the IV estimator was optimal for estimating CACE since it was always unbiased regardless of any other factor related to compliance. When compliance is partial, the Per-Protocol estimator was preferable for estimating CACE in the presence of never-takers only, since it provided less biased results than the IV one. In the presence of never-takers and always-takers, when non-compliance depended on the potential outcomes of participants under usual care, the Per-Protocol estimator was preferable when participants with good or bad conditions would never receive the treatment, while the IV estimator was preferable when the non-random scenarios affected simultaneously participants with good and bad conditions. Finally, when participants with good or bad conditions would always receive the treatment, both estimators can be employed as they provide similar bias. 

Regarding the ATE, our results generally coincide with the findings in \citep{BangDavis2007} and \citep{Yeetal.2014}, with the addition that we considered two additional non-compliant scenarios for the partial compliance case. Shortly, the Per-Protocol estimator was always unbiased for the ATE in the presence of never-takers but it was biased in the presence of never-takers and always-takers except when non-compliance was random, for both binary and partial compliance. On the other hand, the IV estimator was generally biased for the ATE for both binary and partial compliance, except when non-compliance was random or in the case of binary compliance when participants with good conditions always received the active treatment and those with bad conditions always received the control treatment or vice versa.

Our simulations cover a wide range of different non-compliant scenarios by varying three key factors related to non-compliance: degree and randomness of non-compliance and type of non-compliers. Our results are backed by the various sensitivity analyses we have performed. Specifically, including all sensitivity and secondary analyses, we generated a total of $368$ different trial scenarios. We expanded on the non-compliant scenarios considered in previous simulation studies and created a complete guide on the optimal analysis for estimating CACE that can be used by fellow statisticians and clinical researchers to decide whether the Per-Protocol or the IV analysis is most suitable to employ for each non-compliant scenarios.

Despite the advantages of our simulation, our results are limited by a number of factors. First, we only considered the case where the outcomes of the trial are continuous. We did not consider cases where the outcome can be survival or discrete. Additionally, we did not consider specific covariates in the simulation. An adjusted Per-Protocol analysis that accounts for imbalanced covariates between the groups could improve estimation. However, we did consider various associations between participants' outcomes and non-compliant behaviors. Furthermore, the non-compliant scenarios we considered are only a subset of all the possible ones that can occur in clinical practice and their findings may not be generalisable to other scenarios. We believe, however, that the scenarios we considered cover a sufficient number of the non-compliant scenarios that are most likely to be observed in real RCTs and can provide valuable insight for the performance of the two methods. Finally, it is important to note that a simulation analysis is an inevitable simplification and may not fully capture the complexity of the real world, as participants might react differently in real-life situations.


\section{Endmatter}
This is a simulation study. All data generating mechanisms and simulations were implemented in the R Programming Software 4.3.1 version and are available to view and reproduce in GitHub using the following URL address \url{https://github.com/theopapazoglou/Estimating-CACE}.


\clearpage
\textbf{Appendix}
\appendix
\begin{table*}[htbp]
\centering
\renewcommand{\arraystretch}{0.7}
\caption {Results when never-takers and always-takers were allowed. Sensitivity Analysis for treatment effect equal to $0$.} 
\begin{adjustbox}{max width=\textwidth}
\begin{tabular}{cccccc|cccccc}
\hline
\multicolumn{6}{c}{Binary Compliance} & \multicolumn{6}{c}{Partial Compliance} \\
\cmidrule(lr){1-6} \cmidrule(lr){7-12}
Scenario & Method & Estimate & Bias & MSE & Coverage(\%) & Scenario & Method & Estimate & Bias & MSE & Coverage(\%) \\
\midrule
Random & CACE & 0 & - & - & - & Random & CACE & 0 & - & - & - \\
& PP & 0 & 0 & 0.01 & 98\% & & PP & 0.01 & -0.01 & 0.011 & 99\% \\
& IV & 0 & 0 & 0.021 & 98\% & & IV & 0.01 & -0.01 & 0.033 & 98\% \\
\midrule
Scenario A & CACE & -0.51 & - & - & - & Scenario A & CACE & 0 & - & - & - \\
& PP & 0.06 & -0.57 & 0.334 & 0\% & & PP & 0.51 & -0.51 & 0.272 & 0\% \\
& IV & -0.51 & 0 & 0.069 & 97\% & & IV & 0.53 & -0.53 & 0.341 & 57\% \\
\midrule
Scenario B & CACE & 0.51 & - & - & - & Scenario B & CACE & 0.01 & - & - & - \\
& PP & -0.05 & 0.56 & 0.329 & 0\% & & PP & -0.5& 0.51 & 0.267 & 1\% \\
& IV & 0.53 & -0.02 & 0.066 & 96\% & & IV & -0.49 & 0.5 & 0.314 & 63\% \\
\midrule
Scenario C & CACE & -0.51 & - & - & - & Scenario C & CACE & 0 & - & - & - \\
& PP & -0.5 & -0.01 & 0.01 & 97\% & & PP & 0.28& -0.28 & 0.089 & 60\% \\
& IV & -0.51 & 0 & 0.043 & 98\% & & IV & -0.52 & 0.52 & 0.378 & 79\% \\
\midrule
Scenario D & CACE & 0.51 & - & - & - & Scenario D & CACE & 0 & - & - & - \\
& PP & 0.51 & 0 & 0.009 & 97\% & & PP & -0.26 & 0.26 & 0.08 & 68\% \\
& IV & 0.52 & -0.01 & 0.045 & 97\% & & IV & 0.54 & -0.54 & 0.405 & 80\% \\
\midrule
Scenario E & CACE & 0 & - & - & - & Scenario E & CACE & 0 & - & - & - \\
& PP & 0.56 & -0.56 & 0.324 & 0\% & & PP & -0.7 & 0.7 & 0.501 & 0\% \\
& IV & -0.03 & 0.03 & 0.232 & 96\% & & IV & 0.4 & -0.4 & 38 & 98\% \\
\midrule
Scenario F & CACE & 0 & - & - & - & Scenario F & CACE & 0 & - & - & - \\
& PP & -0.55 & 0.55 & 0.322 & 0\% & & PP & 0.71 & -0.71 & 0.505 & 0\% \\
& IV & 0.05 & -0.05 & 0.227 & 98\% & & IV & -0.06 & 0.06 & 0.422 & 97\% \\
\midrule
Scenario G & CACE & - & - & - & - & Scenario G & CACE & 0.06 & - & - & - \\
& PP & - & - & - & - & & PP & 0.93 & -0.87 & 0.765 & 0\% \\
& IV & - & - & - & - & & IV & 0.22 & -0.16 & 99 & 97\% \\
\midrule
Scenario H & CACE & - & - & - & - & Scenario H & CACE & -0.05 & - & - & - \\
& PP & - & - & - & - & & PP & -0.92 & 0.87 & 0.767 & 0\% \\
& IV & - & - & - & - & & IV & 0.14 & -0.19 & 205 & 97\% \\
\hline
\hline
\end{tabular}
\end{adjustbox}
\begin{tabbing}
CACE: True Effect in Compliers, PP: Per-Protocol, IV: Instrumental Variables \\
\end{tabbing}
\end{table*}

\begin{table*}
\centering
\renewcommand{\arraystretch}{0.5}
\caption{Results when only Never-Takers were allowed. Sensitivity Analysis for Treatment Effect equal to $0$.}
\begin{adjustbox}{max width=\textwidth}
\begin{tabular}{cccccc|cccccc}
\toprule
\multicolumn{6}{c}{Binary Compliance} & \multicolumn{6}{c}{Partial Compliance} \\
\cmidrule(lr){1-6} \cmidrule(lr){7-12}
Scenario & Method & Estimate & Bias & MSE & Coverage(\%) & Scenario & Method & Estimate & Bias & MSE & Coverage(\%) \\
\midrule
Random & CACE & 0 & - & - & - & Random & CACE & 0 & - & - & - \\
& PP & 0 & 0 & 0.009 & 99\% & & PP & 0 & 0 & 0.009 & 99\% \\
& IV & 0 & 0 & 0.022 & 98\% & & IV & 0 & 0 & 0.012 & 98\% \\
\midrule
Scenario A & CACE & -0.51 & - & - & - & Scenario A & CACE & -0.13 & - & - & - \\
& PP & 0 & -0.51 & 0.267 & 2\% & & PP & -0.01 & -0.12 & 0.029 & 93\% \\
& IV & -0.51 & 0 & 0.063 & 97\% & & IV & -0.51 & 0.38 & 0.176 & 66\% \\
\midrule
Scenario B & CACE & 0.51 & - & - & - & Scenario B & CACE & 0.13 & - & - & - \\
& PP & 0 & 0.51 & 0.269 & 1\% & & PP & 0 & 0.13 & 0.03 & 93\% \\
& IV & 0.52 & -0.01 & 0.063 & 97\% & & IV & 0.51 & -0.38 & 0.179 & 65\% \\
\midrule
Scenario G & CACE & - & - & - & - & Scenario G & CACE & -0.23 & - & - & - \\
& PP & - & - & - & - & & PP & 0.01 & -0.24 & 0.068 & 62\% \\
& IV & - & - & - & - & & IV & -0.63 & 0.4 & 0.19 & 49\% \\
\midrule
Scenario H & CACE & - & - & - & - & Scenario H & CACE & 0.23 & - & - & - \\
& PP & - & - & - & - & & PP & -0.01 & 0.24 & 0.067 & 63\% \\
& IV & - & - & - & - & & IV & 0.64 & -0.41 & 0.195 & 52\% \\
\bottomrule
\end{tabular}
\end{adjustbox}
\begin{tabbing}
CACE: True Effect in Compliers, PP: Per-Protocol, IV: Instrumental Variables \\
\end{tabbing}
\end{table*}

\begin{table*}
\centering
\renewcommand{\arraystretch}{0.5}
\caption {Results when Never-Takers and Always-Takers were allowed. Sensitivity Analysis for Mild Non-Random Scenarios.} 
\begin{adjustbox}{max width=\textwidth}
\begin{tabular}{cccccc|cccccc}
\toprule
\multicolumn{6}{c}{Binary Compliance} & \multicolumn{6}{c}{Partial Compliance} \\
\cmidrule(lr){1-6} \cmidrule(lr){7-12}
Scenario & Method & Estimate & Bias & MSE & Coverage(\%) & Scenario & Method & Estimate & Bias & MSE & Coverage(\%) \\
\midrule
Scenario A & CACE & -5.29 & - & - & - & Scenario A & CACE & -4.99 & - & - & - \\
& PP & -4.96 & -0.33 & 0.12 & 22\% & & PP & -4.79 & -0.2 & 0.053 & 84\% \\
& IV & -5.29 & 0 & 0.041 & 97\% & & IV & -5.31 & 0.32 & 0.166 & 89\% \\
\midrule
Scenario B & CACE & -4.72 & - & - & - & Scenario B & CACE & -4.99 & - & - & - \\
& PP & -5.04 & 0.32 & 0.117 & 21\% & & PP & -5.2 & 0.21 & 0.055 & 82\% \\
& IV & -4.71 & -0.01 & 0.042 & 97\% & & IV & -4.69 & -0.3 & 0.156 & 91\% \\
\midrule
Scenario C & CACE & -5.29 & - & - & - & Scenario C & CACE & -4.99 & - & - & - \\
& PP & -5.29 & 0 & 0.01 & 98\% & & PP & -5.29 & 0.3 & 0.099 & 38\% \\
& IV & -5.29 & 0 & 0.03 & 98\% & & IV & -5.3 & 0.31 & 0.142 & 77\% \\
\midrule
Scenario D & CACE & -4.72 & - & - & - & Scenario D & CACE & -4.99 & - & - & - \\
& PP & -4.71 & -0.01 & 0.008 & 98\% & & PP & -4.72 & -0.27 & 0.09 & 45\% \\
& IV & -4.71 & -0.01 & 0.028 & 98\% & & IV & -4.72 & -0.27 & 0.123 & 85\% \\
\midrule
Scenario E & CACE & -5 & - & - & - & Scenario E & CACE & -5 & - & - & - \\
& PP & -4.67 & -0.33 & 0.117 & 18\% & & PP & -4.52 & -0.48 & 0.24 & 3\% \\
& IV & -5 & 0 & 0.054 & 98\% & & IV & -5.03 & 0.03 & 0.097 & 97\% \\
\midrule
Scenario F & CACE & -5 & - & - & - & Scenario F & CACE & -4.99 & - & - & - \\
& PP & -5.33 & 0.33 & 0.121 & 21\% & & PP & -5.48 & 0.49 & 0.253 & 3\% \\
& IV & -4.99 & -0.01 & 0.056 & 97\% & & IV & -5 & 0.01 & 0.096 & 97\% \\
\midrule
Scenario G & CACE & - & - & - & - & Scenario G & CACE & -4.87 & - & - & - \\
& PP & - & - & - & - & & PP & -4.31 & -0.56 & 0.327 & 2\% \\
& IV & - & - & - & - & & IV & -4.84 & -0.03 & 0.155 & 96\% \\
\midrule
Scenario H & CACE & - & - & - & - & Scenario H & CACE & -5.12 & - & - & - \\
& PP & - & - & - & - & & PP & -5.7 & 0.58 & 0.347 & 0\% \\
& IV & - & - & - & - & & IV & -5.2 & 0.08 & 0.162 & 95\% \\
\bottomrule
\end{tabular}
\end{adjustbox}
\begin{tabbing}
CACE: True Effect in Compliers, PP: Per-Protocol, IV: Instrumental Variables \\
\end{tabbing}
\end{table*}

\begin{table*}
\centering
\renewcommand{\arraystretch}{0.5}
\caption{Results when only Never-Takers were allowed. Sensitivity Analysis for Mild Non-Random Scenarios.}
\begin{adjustbox}{max width=\textwidth}
\begin{tabular}{cccccc|cccccc}
\toprule
\multicolumn{6}{c}{Binary Compliance} & \multicolumn{6}{c}{Partial Compliance} \\
\cmidrule(lr){1-6} \cmidrule(lr){7-12}
Scenario & Method & Estimate & Bias & MSE & Coverage(\%) & Scenario & Method & Estimate & Bias & MSE & Coverage(\%) \\
\midrule
Scenario A & CACE & -5.29 & - & - & - & Scenario A & CACE & -5.08 & - & - & - \\
& PP & -5 & -0.29 & 0.094 & 36\% & & PP & -4.99 & -0.09 & 0.019 & 96\% \\
& IV & -5.29 & 0 & 0.037 & 96\% & & IV & -5.29 & 0.21 & 0.062 & 87\% \\
\midrule
Scenario B & CACE & -4.72 & - & - & - & Scenario B & CACE & -4.91 & - & - & - \\
& PP & -5 & 0.28 & 0.091 & 39\% & & PP & -4.99 & 0.08 & 0.017 & 97\% \\
& IV & -4.72 & 0 & 0.039 & 98\% & & IV & -4.71 & -0.2 & 0.062 & 87\% \\
\midrule
Scenario G & CACE & - & - & - & - & Scenario G & CACE & -5.12 & - & - & - \\
& PP & - & - & - & - & & PP & -4.99 & -0.13 & 0.03 & 89\% \\
& IV & - & - & - & - & & IV & -5.35 & 0.23 & 0.072 & 80\% \\
\midrule
Scenario H & CACE & - & - & - & - & Scenario H & CACE & -4.87 & - & - & - \\
& PP & - & - & - & - & & PP & -4.99& 0.12 & 0.028 & 88\% \\
& IV & - & - & - & - & & IV & -4.64 & -0.23 & 0.071 & 82\% \\
\bottomrule
\end{tabular}
\end{adjustbox}
\begin{tabbing}
CACE: True Effect in Compliers, PP: Per-Protocol, IV: Instrumental Variables \\
\end{tabbing}
\end{table*}

\begin{table*}
\centering
\renewcommand{\arraystretch}{0.5}
\caption {Results when Never-Takers and Always-Takers were allowed. Sensitivity Analysis for the proportion of Non-Compliers.} 
\begin{adjustbox}{max width=\textwidth}
\begin{tabular}{cccccc|cccccc}
\toprule
\multicolumn{6}{c}{Binary Compliance} & \multicolumn{6}{c}{Partial Compliance} \\
\cmidrule(lr){1-6} \cmidrule(lr){7-12}
Scenario & Method & Estimate & Bias & MSE & Coverage(\%) & Scenario & Method & Estimate & Bias & MSE & Coverage(\%) \\
\midrule
Random & CACE & -4.99 & - & - & - & Random & CACE & -5 & - & - & - \\
& PP & -4.99 & 0 & 0.015 & 96\% & & PP & -5 & 0 & 0.012 & 99\% \\
& IV & -4.98 & -0.01 & 0.066 & 96\% & & IV & -5.02 & 0.02 & 0.11 & 98\% \\
\midrule
Scenario A & CACE & -5.5 & - & - & - & Scenario A & CACE & -5 & - & - & - \\
& PP & -4.91 & -0.59 & 0.37 & 1\% & & PP & -4.81 & -0.19 & 0.053 & 84\% \\
& IV & -5.52 & 0.02 & 0.188 & 96\% & & IV & -5.6 & 0.6 & 0.722 & 92\% \\
\midrule
Scenario B & CACE & -4.49 & - & - & - & Scenario B & CACE & -5 & - & - & - \\
& PP & -5.08 & 0.59 & 0.367 & 1\% & & PP & -5.19 & 0.19 & 0.053 & 88\% \\
& IV & -4.47 & -0.02 & 0.176 & 96\% & & IV & -4.5 & -0.5 & 0.549 & 95\% \\
\midrule
Scenario C & CACE & -5.5 & - & - & - & Scenario C & CACE & -5 & - & - & - \\
& PP & -5.5 & 0 & 0.016 & 92\% & & PP & -5.51 & 0.51 & 0.275 & 1\% \\
& IV & -5.48 & -0.02 & 0.12 & 98\% & & IV & -5.5 & 0.5 & 0.454 & 78\% \\
\midrule
Scenario D & CACE & -4.49 & - & - & - & Scenario D & CACE & -4.99 & - & - & - \\
& PP & -4.49 & 0 & 0.016 & 90\% & & PP & -4.49 & -0.5 & 0.266 & 0\% \\
& IV & -4.48 & -0.01 & 0.116 & 98\% & & IV & -4.49 & -0.5 & 0.501 & 82\% \\
\midrule
Scenario E & CACE & -4.99 & - & - & - & Scenario E & CACE & -5 & - & - & - \\
& PP & -4.43 & -0.56 & 0.344 & 2\% & & PP & -4.35 & -0.65 & 0.433 & 0\% \\
& IV & -5.12 & 0.13 & 1.25 & 97\% & & IV & -5.23 & 0.23 & 35 & 98\% \\
\midrule
Scenario F & CACE & -4.99 & - & - & - & Scenario F & CACE & -5 & - & - & - \\
& PP & -5.57 & 0.58 & 0.351 & 1\% & & PP & -5.66 & 0.66 & 0.441 & 0\% \\
& IV & -5.05 & 0.06 & 13 & 97\% & & IV & -4.96 & -0.04 & 6 & 97\% \\
\midrule
Scenario G & CACE & - & - & - & - & Scenario G & CACE & -4.94 & - & - & - \\
& PP & - & - & - & - & & PP & -4.11 & -0.83 & 0.712 & 0\% \\
& IV & - & - & - & - & & IV & -4.31 & -0.63 & 118 & 98\% \\
\midrule
Scenario H & CACE & - & - & - & - & Scenario H & CACE & -5.06 & - & - & - \\
& PP & - & - & - & - & & PP & -5.9 & 0.84 & 0.718 & 0\% \\
& IV & - & - & - & - & & IV & -5.24 & 0.18 & 57 & 97\% \\
\bottomrule
\end{tabular}
\end{adjustbox}
\begin{tabbing}
CACE: True Effect in Compliers, PP: Per-Protocol, IV: Instrumental Variables \\
\end{tabbing}
\end{table*}

\begin{table*}
\centering
\renewcommand{\arraystretch}{0.5}
\caption{Results when only Never-Takers were allowed. Sensitivity Analysis for the proportion of Non-Compliers.}
\begin{adjustbox}{max width=\textwidth}
\begin{tabular}{cccccc|cccccc}
\toprule
\multicolumn{6}{c}{Binary Compliance} & \multicolumn{6}{c}{Partial Compliance} \\
\cmidrule(lr){1-6} \cmidrule(lr){7-12}
Scenario & Method & Estimate & Bias & MSE & Coverage(\%) & Scenario & Method & Estimate & Bias & MSE & Coverage(\%) \\
\midrule
Random & CACE & -4.99 & - & - & - & Random & CACE & -5 & - & - & - \\
& PP & -4.99 & 0 & 0.015 & 96\% & & PP & -5 & 0 & 0.011 & 99\% \\
& IV & -4.98 & -0.01 & 0.06 & 97\% & & IV & -5.01 & 0.01 & 0.023 & 98\% \\
\midrule
Scenario A & CACE & -5.5 & - & - & - & Scenario A & CACE & -5.11 & - & - & - \\
& PP & -4.99 & -0.51 & 0.281 & 8\% & & PP & -5 & -0.11 & 0.029 & 94\% \\
& IV & -5.51 & 0.01 & 0.152 & 95\% & & IV & -5.53 & 0.42 & 0.247 & 75\% \\
\midrule
Scenario B & CACE & -4.49 & - & - & - & Scenario B & CACE & -4.89 & - & - & - \\
& PP & -4.99 & 0.5 & 0.277 & 7\% & & PP & -5 & 0.11 & 0.03 & 92\% \\
& IV & -4.48 & -0.01 & 0.152 & 96\% & & IV & -4.49 & -0.4 & 0.222 & 80\% \\
\midrule
Scenario G & CACE & - & - & - & - & Scenario G & CACE & -5.24 & - & - & - \\
& PP & - & - & - & - & & PP & -5 & -0.24 & 0.069 & 64\% \\
& IV & - & - & - & - & & IV & -5.71 & 0.47 & 0.26 & 42\% \\
\midrule
Scenario H & CACE & - & - & - & - & Scenario H & CACE & -4.76 & - & - & - \\
& PP & - & - & - & - & & PP & -5.01 & 0.25 & 0.073 & 63\% \\
& IV & - & - & - & - & & IV & -4.31 & -0.45 & 0.239 & 45\% \\
\bottomrule
\end{tabular}
\end{adjustbox}
\begin{tabbing}
CACE: True Effect in Compliers, PP: Per-Protocol, IV: Instrumental Variables \\
\end{tabbing}
\end{table*}

\begin{table*}
\centering
\renewcommand{\arraystretch}{0.5}
\caption {Results when Never-Takers and Always-Takers were allowed. Sensitivity Analysis for different compliance thresholds in Partial Compliance.} 
\begin{adjustbox}{max width=\textwidth}
\begin{tabular}{cc|cccc|cccc|cccc}
\toprule
\multicolumn{2}{c}{} & \multicolumn{4}{c}{CASE II} & \multicolumn{4}{c}{CASE III} & \multicolumn{4}{c}{CASE IV}\\
\cmidrule(lr){3-6} \cmidrule(lr){7-10} \cmidrule(lr){11-14}
Scenario & Method & Estimate & Bias & MSE & Coverage(\%) & Estimate & Bias & MSE & Coverage(\%) & Estimate & Bias & MSE & Coverage(\%) \\
\midrule
Random & CACE & -5 & - & - & - & -5 & - & - & - & -5 & - & - & - \\
& PP & -4.65 & -0.35 & 0.134 & 31\% & -4.38 & -0.62 & 0.391 & 0\% & -4.03 & -0.97 & 0.946 & 0\% \\
& IV & -5.01 & 0.01 & 0.035 & 97\% & -5.01 & 0.01 & 0.037 & 97\% & -5.01 & 0.01 & 0.038 & 97\% \\
\midrule
Scenario A & CACE & -5.06 & - & - & - & -5.08 & - & - & - & -5.11 & - & - & - \\
& PP & -4.34 & -0.72 & 0.529 & 0\% & -4.05 & -1.03 & 1.094 & 0\% & -3.65 & -1.45 & 2.115 & 0\% \\
& IV & -5.55 & 0.49 & 0.358 & 85\% & -5.55 & 0.47 & 0.333 & 88\% & -5.55 & 0.44 & 0.313 & 90\% \\
\midrule
Scenario B & CACE & -4.93 & - & - & - & -4.91 & - & - & - & -4.89 & - & - & - \\
& PP & -4.95 & 0.02 & 0.014 & 98\% & -4.72 & -0.19 & 0.051 & 83\% & -4.4 & -0.49 & 0.253 & 14\% \\
& IV & -4.48 & -0.45 & 0.314 & 87\% & -4.48 & -0.43 & 0.293 & 89\% & -4.48 & -0.41 & 0.276 & 90\% \\
\midrule
Scenario C & CACE & -5.06 & - & - & - & -5.08 & - & - & - & -5.11 & - & - & - \\
& PP & -5.27 & 0.21 & 0.06 & 63\% & -5.07& -0.01 & 0.017& 96\% & -4.8 & -0.31 & 0.116 & 46\% \\
& IV & -5.53 & 0.47 & 0.298 & 68\% & -5.53& 0.45 & 0.275 & 71\% & -5.53 & 0.42 & 0.257 & 74\% \\
\midrule
Scenario D & CACE & -4.93 & - & - & - & -4.91 & - & - & - & -4.89 & - & - & - \\
& PP & -4.3 & -0.63 & 0.413 & 0\% & -4.14& -0.77 & -0.601 & 0\% & -3.91 & -0.98 & 0.959 & 0\% \\
& IV & -4.5 & -0.43 & 0.259 & 72\% & -4.5& -0.41 & 0.239 & 76\% & -4.5 & -0.39 & 0.223 & 78\% \\
\midrule
Scenario E & CACE & -5 & - & - & - & -5 & - & - & - & -5 & - & - & -\\
& PP & -4.13 & -0.87 & 0.757 & 0\% & -3.99& -1.01 & 1.02 & 0\% & -3.79 & -1.21 & 1.48 & 0\% \\
& IV & -5.12 & 0.12 & 0.438 & 97\% & -5.12& 0.12 & 0.438 & 97\% & -5.12 & 0.12 & 0.438 & 97\% \\
\midrule
Scenario F & CACE & -5 & - & - & - & -5 & - & - & - & -5 & - & - & - \\
& PP & -5.54 & 0.54 & 0.313 & 0\% & -5.4& 0.4 & 0.183 & 13\% & -5.2 & 0.2 & 0.062 & 73\% \\
& IV & -4.95 & -0.05 & 0.529 & 96\% & -4.95& -0.05 & 0.53 & 96\% & -4.95 & -0.05 & 0.532 & 96\% \\
\midrule
Scenario G & CACE & -4.97 & - & - & - & -4.98 & - & - & - & -4.98 & - & - & -\\
& PP & -3.79 & -1.18 & 1.394 & 0\% & -3.49& -1.49 & 2.213 & 0\% & -3.12 & -1.86 & 3.458 & 0\% \\
& IV & -5.16 & 0.19 & 21 & 96\% & -5.16& 0.18 & 21 & 96\% & -5.16 & 0.18 & 21 & 96\% \\
\midrule
Scenario H & CACE & -5.03 & - & - & - & -5.02 & - & - & - & -5.02 & - & - & - \\
& PP & -5.68 & 0.65 & 0.438 & 0\% & -5.4& 0.38 & 0.163 & 24\% & -5.02 & 0 & 0.028 & 94\% \\
& IV & -5.23 & 0.2 & 34 & 95\% & -5.23& 0.21 & 34 & 94\% & -5.23 & 0.21 & 34 & 95\% \\
\bottomrule
\end{tabular}
\end{adjustbox}
\begin{tabbing}
CACE: True Effect in Compliers, PP: Per-Protocol, IV: Instrumental Variables \\
\end{tabbing}
\end{table*}

\begin{table*}
\centering
\renewcommand{\arraystretch}{0.5}
\caption {Results when only Never-Takers were allowed. Sensitivity Analysis for different compliance thresholds in Partial Compliance.} 
\begin{adjustbox}{max width=\textwidth}
\begin{tabular}{cc|cccc|cccc|cccc}
\toprule
\multicolumn{2}{c}{} & \multicolumn{4}{c}{CASE II} & \multicolumn{4}{c}{CASE III} & \multicolumn{4}{c}{CASE IV}\\
\cmidrule(lr){3-6} \cmidrule(lr){7-10} \cmidrule(lr){11-14}
Scenario & Method & Estimate & Bias & MSE & Coverage(\%) & Estimate & Bias & MSE & Coverage(\%) & Estimate & Bias & MSE & Coverage(\%) \\
\midrule
Random & CACE & -4.99 & - & - & - & -4.99 & - & - & - & -4.99 & - & - & - \\
& PP & -4.64 & -0.35 & 0.138 & 11\% & -4.38 & -0.61 & 0.395 & 0\% & -4.03 & -0.96 & 0.951 & 0\% \\
& IV & -4.99 & 0 & 0.011 & 99\% & -4.99 & 0 & 0.012 & 98\% & -4.99 & 0 & 0.013 & 98\% \\
\midrule
Scenario A & CACE & -5.16 & - & - & - & -5.18 & - & - & - & -5.19 & - & - & - \\
& PP & -4.6 & -0.56 & 0.33 & 1\% & -4.31 & -0.87 & 0.763 & 0\% & -3.93 & -1.26 & 1.624 & 0\% \\
& IV & -5.52 & 0.36 & 0.161 & 70\% & -5.52 & 0.34 & 0.152 & 74\% & -5.52 & 0.33 & 0.144 & 76\% \\
\midrule
Scenario B & CACE & -4.83 & - & - & - & -4.81 & - & - & - & -4.8 & - & - & - \\
& PP & -4.67 & -0.16 & 0.037 & 86\% & -4.44 & -0.37 & 0.157 & 21\% & -4.13 & -0.67 & 0.472 & 0\% \\
& IV & -4.48 & -0.35 & 0.158 & 73\% & -4.48 & -0.33 & 0.148 & 75\% & -4.48 & -0.32 & 0.14 & 77\% \\
\midrule
Scenario G & CACE & -5.23 & - & - & - & -5.22 & - & - & - & -5.21 & - & - & -\\
& PP & -4.71 & -0.52 & 0.272 & 1\% & -4.41& -0.81 & 0.66 & 0\% & -4.04 & -1.17 & 1.385 & 0\% \\
& IV & -5.64 & 0.41 & 0.206 & 48\% & -5.64& 0.42 & 0.21 & 46\% & -5.64 & 0.43 & 0.221 & 43\% \\
\midrule
Scenario H & CACE & -4.77 & - & - & - & -4.77 & - & - & - & -4.78 & - & - & - \\
& PP & -4.74 & -0.03 & 0.013 & 98\% & -4.45& -0.32 & 0.116 & 39\% & -4.08 & -0.7 & 0.518 & 0\% \\
& IV & -4.36 & -0.41 & 0.2 & 51\% & -4.36& -0.41 & 0.204 & 50\% & -4.36 & -0.42 & 0.214 & 46\% \\
\bottomrule
\end{tabular}
\end{adjustbox}
\begin{tabbing}
CACE: True Effect in Compliers, PP: Per-Protocol, IV: Instrumental Variables \\
\end{tabbing}
\end{table*}



\end{document}